\documentclass[twocolumn,aps,pre,showpacs,amsmath,amsfonts,amssymb,floatfix]{revtex4}

\usepackage{morefloats}
\usepackage{amsmath}
\usepackage{graphicx}
\usepackage{dcolumn}
\usepackage{bm}
\usepackage{color}
\usepackage{amssymb}
\definecolor{orange}{rgb}{1,0.5,0}
\begin{document}
\title{Cluster Synchronization in Multiplex Networks}
\author{Sarika Jalan$^{1,2}$ and Aradhana Singh$^1$}
\affiliation{$^1$Complex Systems Lab, Discipline of Physics, Indian Institute of Technology Indore, Indore-452017}
\affiliation{$^2$Centre for Bio-Science and Bio-Medical Engineering, Indian Institute of Technology Indore, Indore-452017}

\begin {abstract}
We study the impact of interaction of nodes in a layer of a multiplex network
 on the dynamical behavior and cluster synchronization
of these nodes in other layers. 
We find that nodes interactions in one layer affects the 
cluster synchronizability of another layer in many different ways. 
While multiplexing with a sparse network enhances the synchronizability 
multiplexing with a dense network suppresses the cluster synchronizability with
the network architecture deciding the impact of the enhancement and 
suppression. Additionally, at weak couplings the enhancement in the cluster 
synchronizability due to multiplexing remains of the driven type, while  
for strong couplings the multiplexing  may lead to a transition to the
self-organized mechanism. 
%In both cases there is change in the cluster patterns.

\end{abstract}
\pacs{05.45.Xt,05.45.Pq}
\maketitle
Synchronization is an universal phenomenon observed in a range of 
the systems \cite{Book_syn}. It is known to be important for proper 
functioning of many complex systems. For example in brain, the phase 
synchronization at  different locations is responsible for action, motion, 
vision and sensation \cite{Cluster_Brain}. Recently, cluster 
synchronization has gained tremendous attention
due to its occurrence and importance in real world systems \cite{Clus_syn_review, SJ_prl2003,Clus_syn2}. The cluster synchronization refers to the case when the nodes of a 
system divide into several synchronized groups so that the nodes in
the same group synchronize with each other while do not synchronize with
the nodes in the different groups. 
So far the studies on the cluster synchronization have mainly focused on 
complex systems being represented as isolated networks
\cite{Nature,PRE,Laser_Kanter,Scholl,Chaos},  
however a complex system may consist of a superposition of a number of 
interacting networks \cite{Boccaletti,pnas,networks}, such as a social system which is composed of different sub-networks consisting family, friends, colleagues, work collaborators and hence forming a multiplex network. 
The multiplex network presents a more realistic representation of 
real world interactions \cite{Boccaletti} leading to 
a spurt in the
activities of modeling real world complex systems.
Most of these studies have 
concentrated on the investigation of various structural properties or emergence of spectral properties \cite{Multiplex_Arenas1, Multiplex_Arenas2, Multiplex_degree}. Few works considering dynamical 
properties of the multiplex networks report that the synchronizability of a 
multiplex network is maximum for the small-world - random regular topology \cite{Syn_multiplex} and multiplexing 
reduces the rate of the global synchronization \cite{Syn_multiplex}.  

In this Rapid communication, we study dynamical behavior of nodes in a 
layer upon multiplexing with another layer. Particularly, we investigate the 
impact of nodes interactions in one layer on the cluster synchronization of the same nodes in the other layer. 
In a realistic situation, the connection density as well as degree distribution of two 
layers can be different, 
for instance in a social system a family network can be denser than a counter friendship network.
Similarly, the friendship network can be denser than a corresponding business network.  
In the present paper, we explore the impact of the network architecture on the cluster synchronizability of another layer. 
 We find that nodes interacting in one layer affect the synchronizability of another layer in many different ways.  While multiplexing with sparse networks enhances the cluster synchronizability of a layer, impact of multiplexing with dense networks depend on the network architecture.
The enhancement in cluster synchronization is referred 
to when number of the nodes forming synchronized clusters increases. 
Furthermore, we report that the change in the density of connections in a 
layer of the multiplex network may bring a change in the mechanism of the 
cluster synchronization in another 
layer. Previous studies on dynamical behavior of isolated networks have identified two different 
mechanisms of cluster synchronization namely, the driven (D) and the self-organized 
(SO) \cite{SJ_prl2003}. The D and SO synchronization correspond to the synchronization due to the inter and intra cluster couplings, respectively.
%
%We find that at strong couplings the enhancement in the connection density of one layer may bring a transition from the D to the SO mechanism of the cluster formation in another layer.

\begin{figure}[b]
\includegraphics[width=0.75\columnwidth]{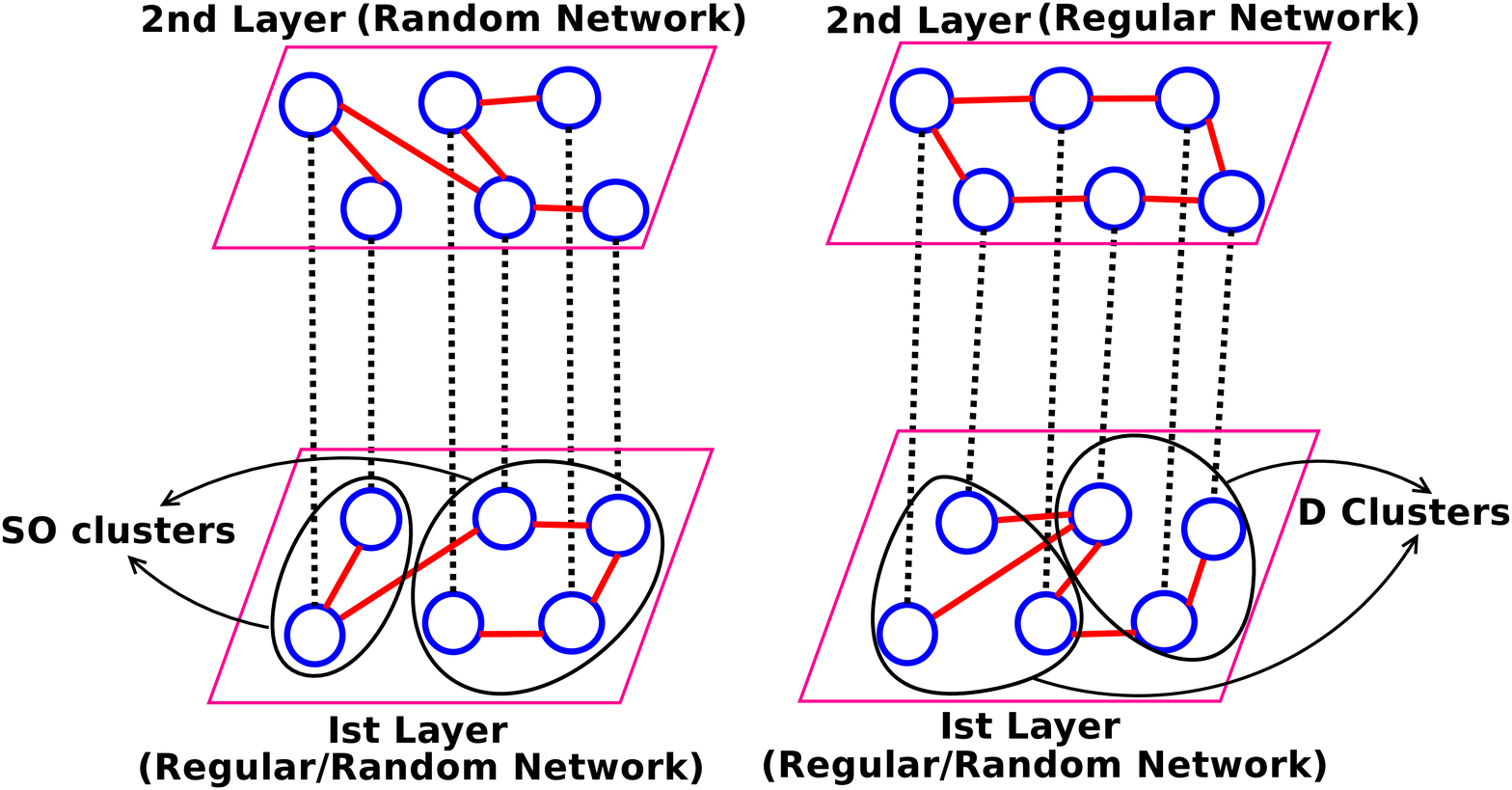}
\caption{Schematic diagram depicting a multiplex network with two layers. The dashed lines indicate 
the inter-layer connections. The density of connections in the different layers can be different and is 
defined as $\langle k_1 \rangle$ for the first layer and $\langle k_2 \rangle$ for the second layer.}
\label{Fig_Multiplex}
\end{figure}
We consider the well known coupled maps model \cite{rev_cml} to investigate
the phase synchronized clusters in the multiplex networks. 
We consider the phase synchronization instead of the complete synchronization as for sparse networks number of nodes 
exhibiting the complete 
synchronization is very less and with an increase in the connection density there is a transition to the globally synchronized state \cite{Boccaletti}, whereas the prime motive of the current work is to study cluster 
synchronization. The phase synchronization reveals interesting cluster patterns as well as 
dependence of mechanism of cluster formation in one layer on the network structure of another layer.
Let each node of the network
be assigned a dynamical variable $x^i, i = 1, 2, \hdots, N$. 
The dynamical evolution is defined by,
\begin{equation}
x_i(t+1) = (1-\varepsilon) f(x_i(t)) + \frac{\varepsilon}{k_i} \sum_{j=1}^N A_{ij} g(x_j(t))
\label{cml}
\end{equation}
Here, $A$ is the adjacency matrix with elements
$A_{ij}$ taking values $1$ and $0$ depending upon whether there is a connection between $i$ and $j$ or not. Degree of a node is given as, $ k_{i}$ = $\sum_{j=1}^{N}A_{ij}$  and $\varepsilon$ is the overall coupling constant and $N$ is total number of nodes in a layer.
The average degree of the different layers  may be different and are indicated as $\langle k_1 \rangle$ and $\langle k_2 \rangle$. 
The function $f(x)$ defines a local nonlinear map, whereas $g(x)$
defines the nature of coupling between the nodes. 
We consider phase synchronization defined as follows \cite{SJ_prl2003}. 
Let $n_i$ and $n_j$ denote the number of times when the 
variables $x_i(t)$ and $x_j(t)$, $t=1, 2, \hdots T$ for the nodes $i$ and $j$ exhibit local minima
during the time interval $T$. Let $n_{ij}$ denotes the number of times these local minima match with 
each other. The phase distance between two nodes $i$ and $j$ is then given as
$d_{ij} = 1 - 2n_{ij}/(n_i + n_j)$.
The nodes $i$ and $j$ are phase synchronized if $d_{ij}=0$. 
All the pairs of nodes in a cluster are phase synchronized \cite{Phase2}.

\begin{figure}[t]
\includegraphics[width=1.3in, height=1.05in]{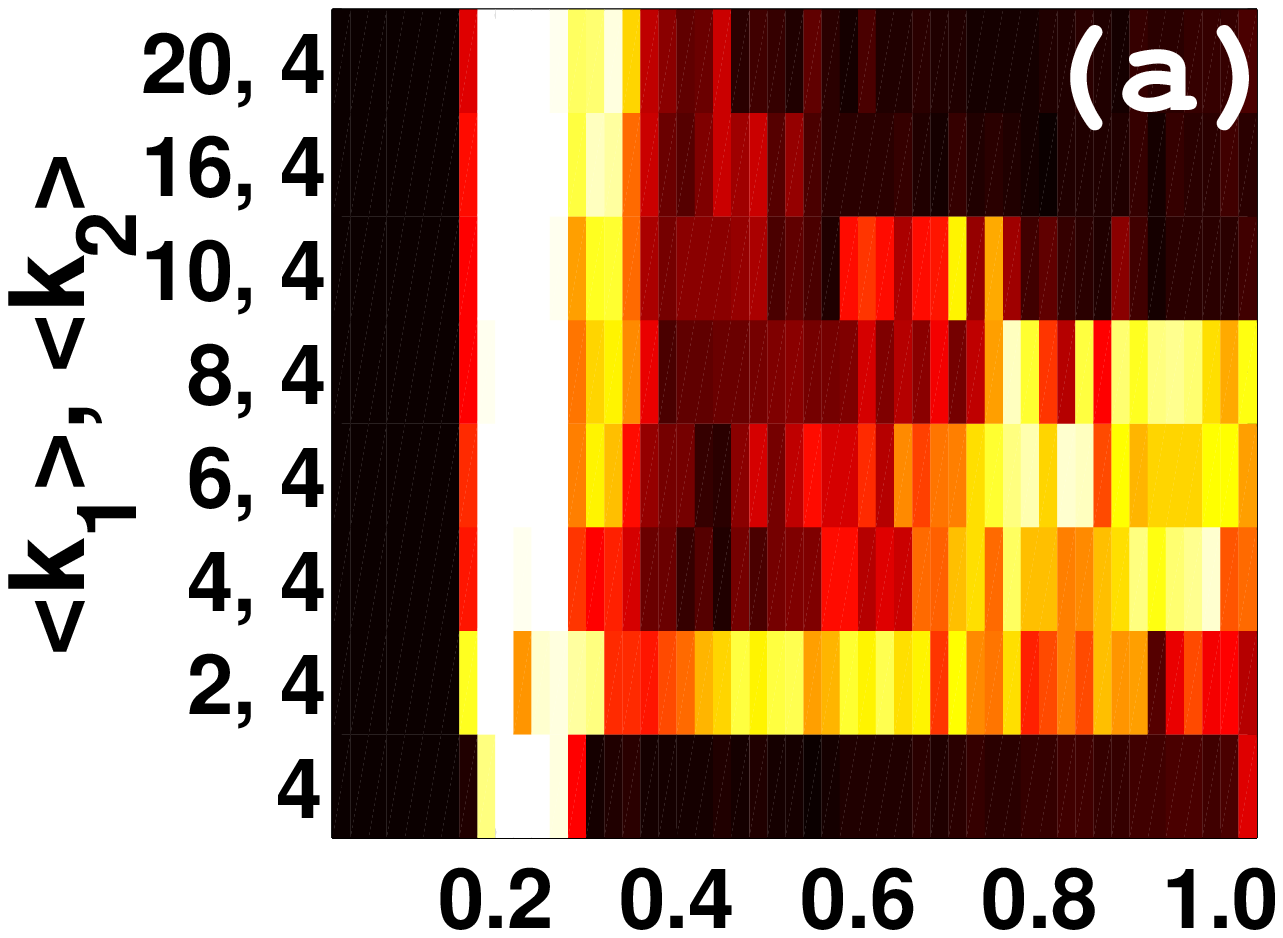}
\includegraphics[width=0.95in, height=1.06in]{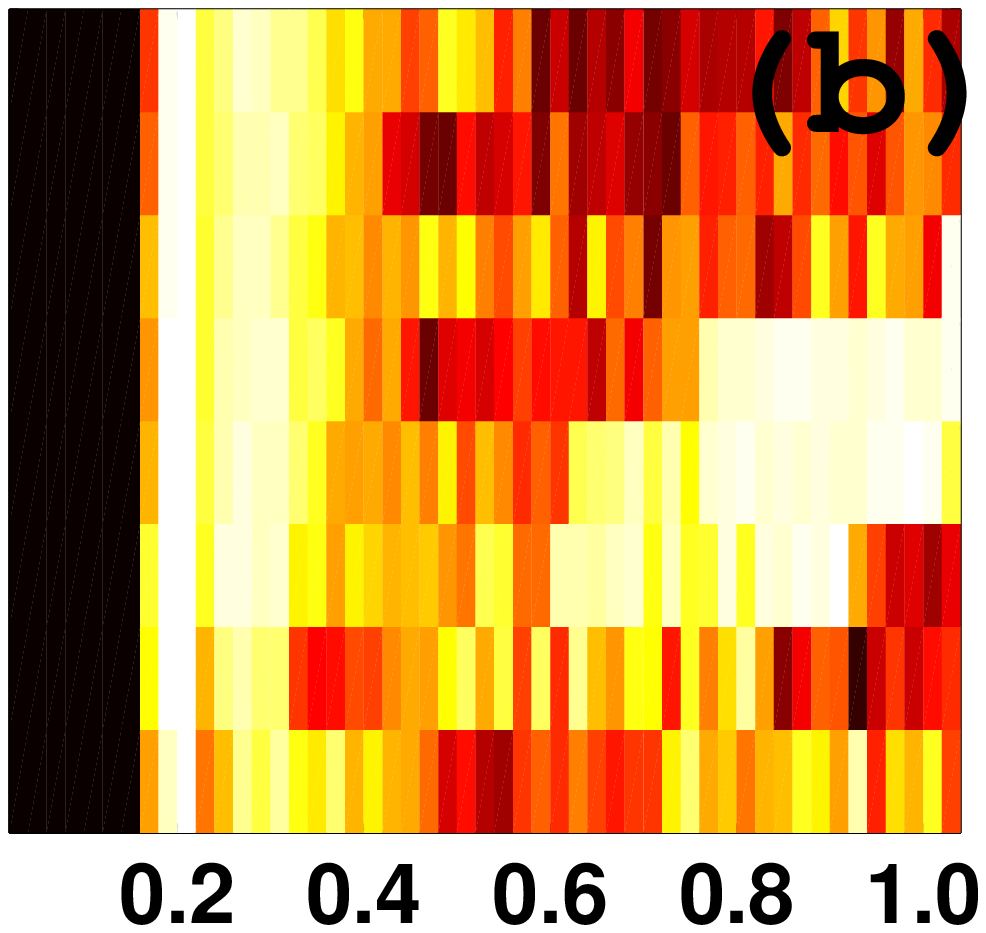}
\includegraphics[width=1.05in, height=1.04in]{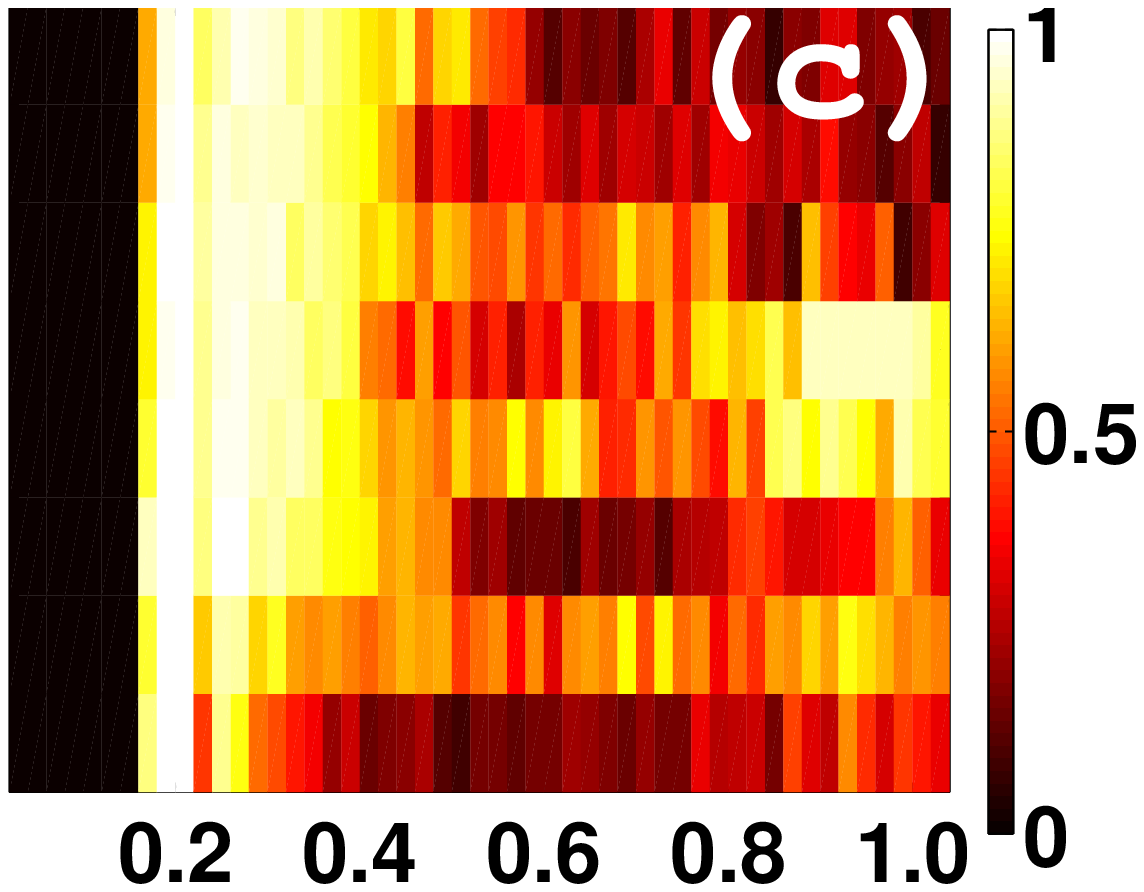}
\includegraphics[width=1.23in, height=1.18in]{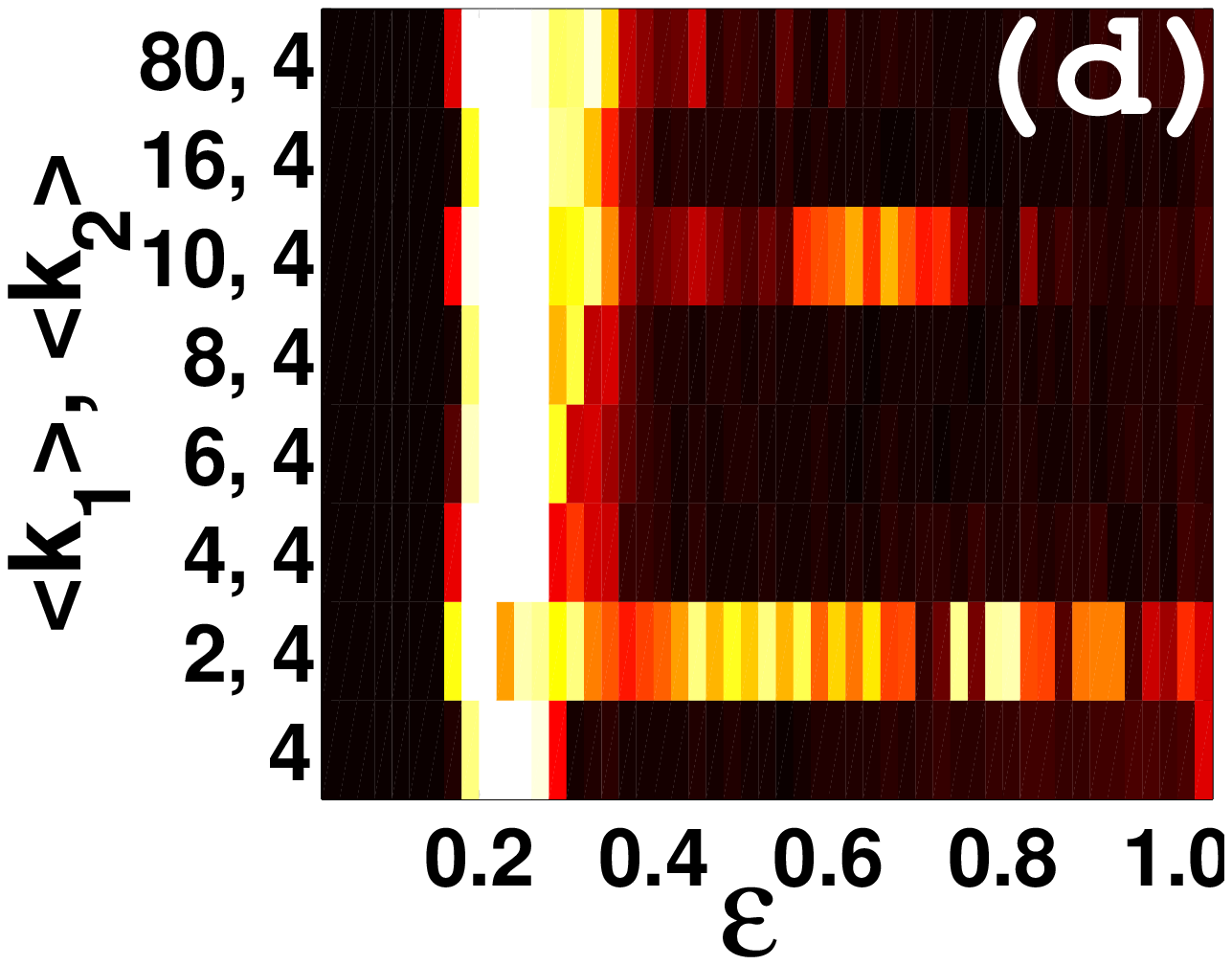}
\includegraphics[width=0.95in, height=1.16in]{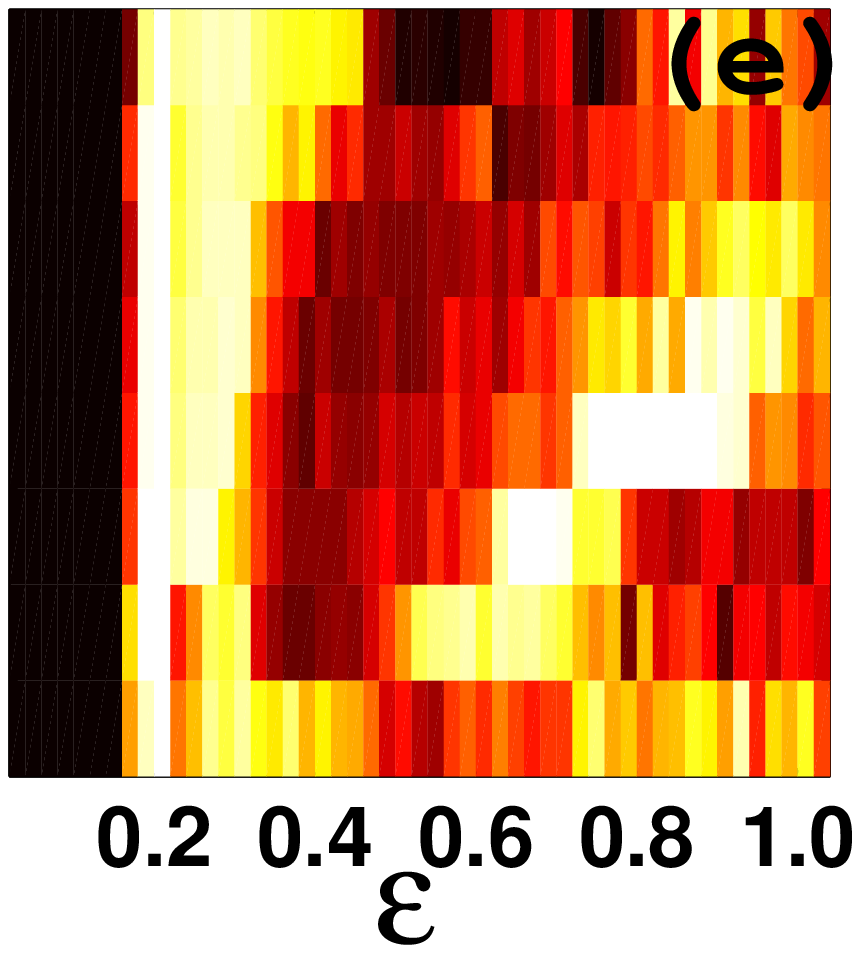}
\includegraphics[width=1.1in, height=1.16in]{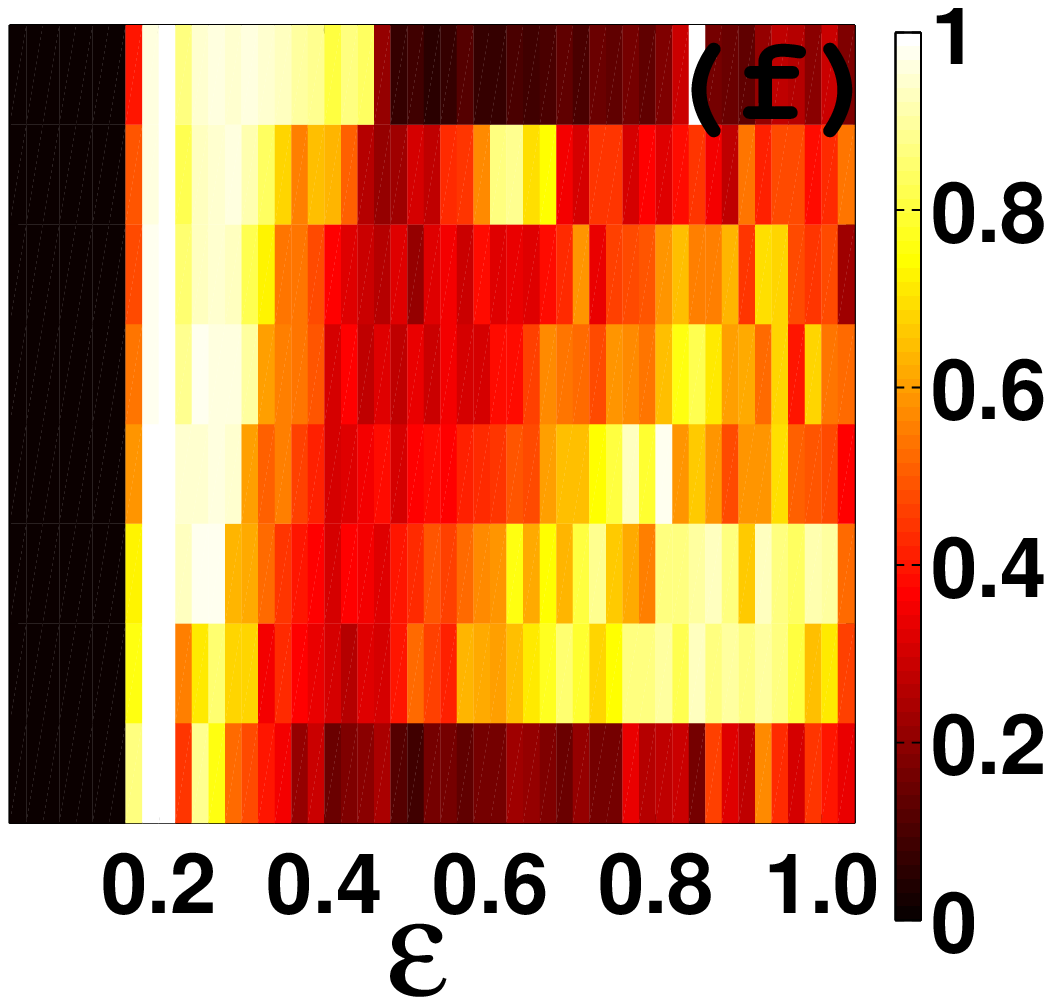}
\caption{Phase diagram depicting the variation of $f_{clus}$ with respect to the $\varepsilon$ and the average degree ($\langle k_2 \rangle$) of a layer with 
1-d lattice (top panel) and  random network (bottom panel) architecture for  
(a) 1-d lattice, (b) SF networks and (c) random networks. The average degree of the second layer ($\langle k_1 \rangle$) remains same for all three cases. For all the layers $N=100$ and phase diagrams are plotted for average over 20 random
realizations of the networks and initial conditions.}
\label{Fig_phase}
\end{figure}
\begin{figure}[t]
\includegraphics[width=0.95\columnwidth]{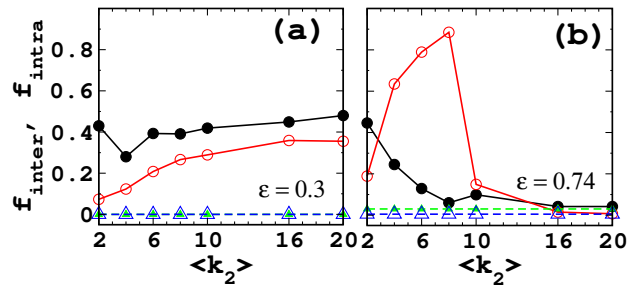}
\caption{Variation of $f_{inter}$ and $f_{intra}$ with $\langle k_2 \rangle$ for isolated 
1-d lattice (close and open triangles) with $N=100$ and $\langle k_1 \rangle = 4$ and after multiplexing with a random network (close and open circles) with various average degrees ($\langle k_2 \rangle$).  All the graphs are plotted for an average over twenty realizations of the initial conditions. Value of $\varepsilon$ are chosen such that they exhibit an enhancement in the D synchronization  and enhancement in the SO synchronization followed by a suppression at the strong couplings with an increase in $\langle k_2 \rangle$. All the graphs are plotted for average over $20$ different realizations of network and initial conditions. }
\label{Fig_finter_NNC}
\end{figure}
\begin{figure}[t]
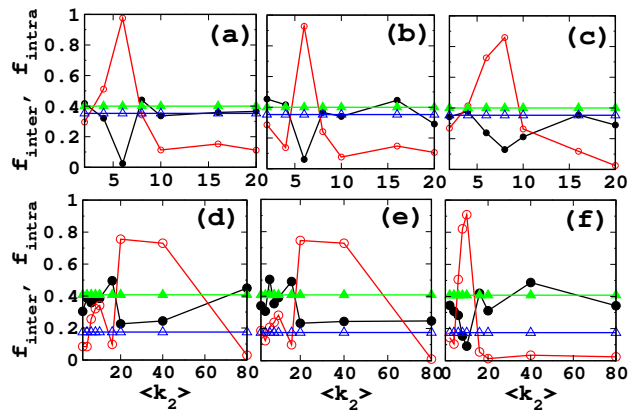

\includegraphics[width=0.95\columnwidth]{Fig4a.eps}
\includegraphics[width=0.95\columnwidth]{Fig4b.eps}
\caption{Variation of $f_{inter}$ and $f_{intra}$ with $\langle k_2 \rangle$ for isolated SF network
(close and open triangles) and for SF network after multiplexing (closed and open circles) with (a) 1-d lattice, (b) SF networks, and (c) random networks at $\varepsilon=0.74$ (top panel) and 
$\varepsilon=1.0$ (bottom panel). All the graphs are plotted for average over $20$ different realizations of network and initial conditions.}
\label{Fig_finter_SF}
\end{figure}
\begin{figure}[t]
\centerline{\includegraphics[width=0.62\columnwidth]{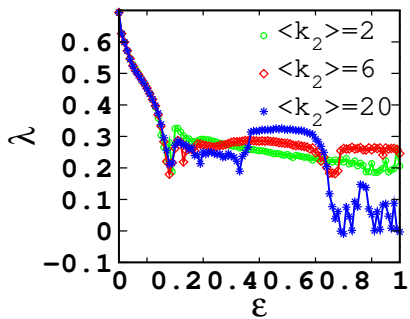}}
\caption{The largest Lyapunov exponent for a multiplex network consisting of two layers, one represented
with the ER random ($\langle k_1 \rangle =4$) network and another with 1-d lattice for various average degree $\langle k_2 \rangle$. Number of nodes in each layer is 
$N=100$.}
\label{Fig_lya}
\end{figure}

We evolve Eq.~\ref{cml} starting from a set of random initial conditions and study the phase 
synchronized clusters after an initial transient. We present detailed results of cluster synchronization for simplest multiplex network consisting of two layers.
First layer can be represented by a regular or a random network, similarly the second layer can also be 
modeled by a regular or random network. 
Here, we present results for all the possible combinations, such as random-random, random-regular, regular-regular and regular-random. 
First, we discuss the cluster synchronizability of a regular network represented by 1-d lattice upon 
multiplexing with 
a ER random network \cite{networks}.
We find that the isolated sparse 1-d lattice at weak couplings leads to the phase synchronized 
clusters with all the nodes participating in the clusters, whereas strong couplings lead to a very few 
nodes forming  clusters (Fig.~\ref{Fig_phase}(a)).  
Multiplexing with a sparse ER network enhances the cluster synchronizability of the 1-d lattice at all the couplings. Multiplexing with a denser ER network 
while enhances the cluster
synchronizability as weak couplings, leaves the cluster
synchronizability unchanged with few nodes keep forming synchronized clusters at the intermediate and strong couplings.
Fig.~\ref{Fig_phase}(a) demonstrates that cluster synchronizability of 1-d lattice enhances at the weak couplings irrespective of the value of $\langle k_2 \rangle$, whereas at the 
intermediate coupling, for $\langle k_2 \rangle = 2$, there is an enhancement in the cluster synchronization, which for the higher values of $\langle k_2 \rangle $ gets vanished. At strong couplings synchronization enhances for $\langle k_2 \rangle \lesssim 8$. 
Additionally, the multiplex network yields the chaotic dynamics for almost all the coupling values for 
the layers being represented by sparse networks. Note that the synchronizability of the second layer, represented as the ER random network, always increases with an increase in the average degree as observed for the isolated networks. 

Next we discuss the cluster synchronizability of the 1-d lattice upon multiplexing with various other
network architectures. At the weak couplings, multiplexing with the 1-d lattice and SF networks lead to an 
enhancement in the cluster synchronizability as observed for the multiplexing with the ER random network. At the strong couplings, there is an enhancement in the synchronization for sparse networks as observed for the multiplexing with the random networks but the connection density for which this enhancement
occurs becomes lower.
For example, a 1-d lattice with $\langle k_1 \rangle = 4$, exhibits 
no enhancement in the cluster synchronization upon multiplexing with the 1-d lattice and SF networks with $\langle k_2 \rangle \gtrsim 4$ and $\langle k_2 \rangle \gtrsim 6$ respectively (Fig.~\ref{Fig_phase}(a)).
Thus the enhancement in the cluster synchronizability of the 1-d lattice is least favorable when it is multiplexed with the 1-d lattice and favorable being
 multiplexed with the random networks.

Further, we study cluster synchronizability of SF networks upon multiplexing with various 
network architecture. The isolated sparse SF networks are known to exhibit a better cluster synchronizability
as compared to the sparse regular or random networks (Fig.~\ref{Fig_phase}(b)). At the 
weak couplings, the multiplexing with another network
only changes the cluster pattern and does not bring any enhancement in the cluster synchronization. For example the isolated SF networks with $\langle k_1 \rangle = 4$ and $N=100$ lead to the 
participation of  about $50\%$ nodes in the cluster formation. After multiplexing, the same fraction of the nodes keep participating in the cluster formation as shown by the reappearance of the grey shade in the Fig.~\ref{Fig_phase}(b).
 At the intermediate and strong couplings there is an enhancement in the synchronization for multiplexing with the sparse networks whereas impact of multiplexing with the dense networks largely depend on the network architecture.
 For example, the cluster synchronizability of SF networks with $\langle k_1 \rangle = 4$ enhances for $\langle k_2 \rangle \lesssim 40$ (Fig.~\ref{Fig_finter_SF}(e) and (f)) upon the multiplexing with 1-d lattice and SF networks. For the higher connection density there is a suppression in the synchronization, while in the case of multiplexing with the random network the enhancement occurs for $\langle k_2 \rangle \lesssim 12$ (Fig.~\ref{Fig_finter_SF}(f)) and with a further increase in the connection density cluster synchronization suppresses. This shows that the cluster synchronizability of the SF network is favorable when it is multiplexed with the 1-d lattice and SF networks.   
Furthermore, multiplexing of ER random networks with different network architectures 
at the weak couplings exhibit the similar behavior as observed for the 1-d 
lattice, whereas the strong couplings lead to the similar behavior as 
discussed for the SF networks.  What follows that multiplexing of random (SF and ER)
networks with 1-d lattice favors more to the cluster synchronizability as compared to multiplexing with the random networks.

%Further, multiplexing of the globally connected network with the random ER, SF
%and 1-d lattice lead to the destruction of the global 
%synchronized state, observed for the isolated network at the intermediate and strong couplings. 
%Additionally multiplexing with the sparse networks lead to a better cluster synchronizability. For example, multiplexing of a globally connected network with a sparse ER 
%random network of $\langle k \rangle \lesssim 8$ reflects a cluster consisting of almost all the nodes, while for 
%$\langle k \rangle = 10 $, less than the 
%$45\%$ of the nodes participate in the cluster formation (Fig.~\ref{Fig_phase_glob}(b)). 
%For $\langle k \rangle > 10$, the synchronization suppresses completely (Fig.~\ref{Fig_phase_glob}). Additionally multiplexing of the globally connected network with the 1-d lattice 
%exhibits a  better cluster synchronizability than multiplexing with random networks. 

In the following, we explore the reasons behind the 
impact of change in the density of connections in 
the second layer on the cluster synchronizability of first layer at strong couplings by using a simple case.
The difference variable, of two nodes in the first layer at $\varepsilon=1$, can be written as,
 \begin{eqnarray}
x_i^1(t+1)-x_j^1(t+1) = \frac{1}{k_i^1 +1} (\sum_{j=1}^N (A_{ij}^1 f(x_i^1(t))))- \nonumber\\
\frac{1}{k_j^1 +1}(A_{ji}^1 f(x_j^1(t))) + (\frac{1}{k_i^1 +1}f(x_i^2(t)- 
\nonumber\\
\frac{1}{k_j^1 +1}f(x_j^2(t)))),
\nonumber\\
\end{eqnarray}
where superscripts 1 and 2 stand for the first and second layer respectively. 
If global synchronization is achieved in the second layer, due to its denseness, in the above variable the coupling term having contribution from the second layer will get cancel out provided  these pair of nodes have same degree ($k_i = k_j$). Consequently the synchronization between two nodes will depend only on the properties of isolated network. For example, the sparse 1-d lattice upon multiplexing with dense random networks exhibits no cluster synchronization as observed for the isolated network (Fig.~\ref{Fig_phase}(a)).  
 
\begin{figure}[t]
\includegraphics[width=0.72\columnwidth]{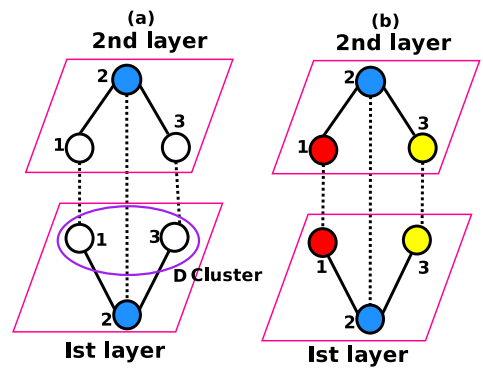}
\caption{Schematic diagrams depicting a multiplex network with three nodes in each layer. (a) formation of D cluster (nodes within the circle)  when node 2 is synchronized with its counter part (denoted with same shade (color), (b) complete suppression in synchronization due to synchronization between the all the nodes (denoted with their counter parts (color)).}
\label{Fig_Three}
\end{figure}

Furthermore, we study the change in the mechanism behind the cluster formation 
due to multiplexing. At weak couplings, the D synchronization 
remains the prime mechanism behind the cluster formation
after multiplexing (Fig.~\ref{Fig_finter_NNC}(a)) as observed for the isolated networks \cite{SJ_pre2005}, while at the intermediate and strong couplings the connection density of the second layer plays an important role. In this coupling regime, multiplexing 
with the sparse networks leads to a change in the mechanism of cluster synchronization(Fig.~\ref{Fig_finter_NNC}(b) and Fig.~\ref{Fig_finter_SF}), whereas upon multiplexing with  the dense networks though there is a suppression in cluster synchronization but the mechanism
behind cluster formation remains same (Fig.~\ref{Fig_finter_NNC}(b)).
In order to understand this impact of the multiplexing on the mechanism of the cluster formation, we investigate cluster synchronizability of a three nodes network. We find that the synchronization among the nodes in the same layer is suppressed due to an enhancement in the synchronization between the nodes of the different layers.  
What follows that the suppression in the SO synchronization at the strong couplings occurs due to the synchronization between nodes which are counter part of each other in different layers, whereas the D synchronization between a pair of node remains unaffected due to the same coupling environment they receive. 
As in Fig.~\ref{Fig_Three}(a), occurrence of synchronization between node 2 in first layer with its counter part in the second layer suppresses the synchronization between nodes 2, 3 and 2, 1 while the nodes 1 and 3 remain synchronized as these nodes still receive a common coupling from node 2.

\begin{figure}[t]
\includegraphics[width=0.75\columnwidth]{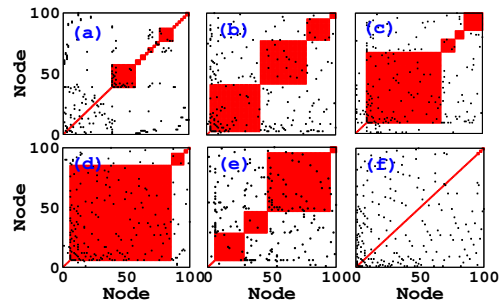}
\caption{Node versus node diagram (a) for the isolated SF network with $N=100$ and $\langle k_2 \rangle =2$, (b), (c), (d), (e) and (f) after multiplexing with a layer represented by ER random network with $k_2 \rangle = 4, 6, 8, 10, 16$ respectively at $\varepsilon=0.8$.}
\label{Fig_Rand_SF_clus}
\end{figure}

%\begin{figure}
%\includegraphics[width=0.75\columnwidth]{Multiplex_clus_NNC.eps}
%\caption{ Node versus node diagram showing the enhancement in the cluster synchronization with multiplexing and suppression of the synchronization when the density of the randomly connected layer increases above some critical value. (a) shows clusters for the isolated NNC network of $N=100$ and $\langle k_1 \rangle =4$, (b), (c), (d), (e) and (f) shows the cluster formation in the same SF network after multiplexing with a randomly connected layer of $N=100$ and $\langle k_2 \rangle = 4, 6, 8, 10 and 16$ respectively.}
%\label{Fig_Rand_1-d_clus}
%\end{figure}
%\begin{figure}[t]
%\includegraphics[width=0.32\columnwidth]{NNC_Glob_clus.eps}
%\includegraphics[width=0.32\columnwidth]{Rand_Glob_clus.eps}
%\caption{Phase diagram depicting the variation of $f_{clus}$ of globally connected network with respect to the $\varepsilon$ and the average degree ($\langle k_2\rangle$)of the a layer with 1-d lattice(a) regular (b) random network architecture. For all the networks $N=100$. The figure is obtained by averaging over 10 initial conditions.}
%\label{Fig_phase_glob}
%\end{figure}

To conclude, we have studied the impact of multiplexing on the cluster synchronizability and mechanism behind the synchronization of a layer in the simplest multiplex network consisting of a network. We find that at weak couplings, the multiplexing enhances the cluster synchronizability, while at the strong couplings this enhancement depends on the architecture as well as the connection density of the another layer.  
The cluster synchronizability of a layer is enhanced when another layer has a moderate connection density. Moreover, multiplexing favors to the enhancement in the cluster synchronization when one of the layer is random and another is regular. The enhancement in the 
cluster synchronization is also associated with the change in the mechanism of the cluster formation. 
The multiplexing primarily influences the synchronization between the nodes which are directly connected while leaving the synchronization between other pairs of nodes
unaffected.

Our work demonstrates that in a multiplex network, the activity in a layer (sub-network) is 
significantly influenced by the structural properties  of another layer (sub-network). If connection density in one layer increases above a certain limit, it may spoil the synchronization
in the another layer. 
The results presented here, about dynamical behavior
of multiplex networks, may provide a guidance for the construction of a 
better model networks with multiplex architecture, such as the airport networks composing different airline companies  \cite{airport_network}.

SJ acknowledges CSIR (25(0205)/12/EMR-II) for the financial 
support. AS thanks complex systems lab members, specially Alok Yadav for useful discussions.


\begin{thebibliography}{99}
\bibitem{network}1-D lattices used in the simulation have circular boundary conditions
with each node having $k$ nearest neighbors. Small-world, scale-free and random networks
are obtained by using SW, BA and ER model respectively
\cite{rev-network}. The complete bipartite networks consist of two sets where each node 
of one set is connected with all the nodes of other set. Results are presented for
both the sets having equal number of nodes.
\bibitem{Phase2} Note that the definition of phase
synchronization used here assign
anti-phase synchronized nodes into two different clusters
due to phase distance being one. However,
this particular situation of nodes being anti-phase synchronized is not so often observed for chaotic evolution.
\end{thebibliography}
\end{document}